\documentclass[doublecol]{epl2}
\usepackage{graphicx}
\usepackage{epsfig}
\usepackage{subeqn}

\newcommand{\bk}{{\bf k}}

\newcommand{\bt}{{\bf t}}

\newcommand{\bx}{{\bf x}}
\newcommand{\by}{{\bf y}}
\newcommand{\bz}{{\bf z}}

\newcommand{\bgamma}{\mbox{\boldmath$\gamma$}}

\newcommand{\bS}{{\bf S}}
\newcommand{\bT}{{\bf T}}

\newcommand{\kbt}{{k_{\rm B}T}}

\date{\today}

\title{Scattering of phonons on two-level systems in disordered crystals}

\author{Drago\c s-Victor Anghel\inst{1} \and Dmitry Churochkin\inst{2}\inst{3}}
\institute{
  \inst{1} Department of Theoretical Physics, Horia Hulubei National Institute for Physics and Nuclear Engineering (IFIN-HH), 407 Atomistilor, Magurele-Bucharest 077125, Romania, EU\\
  \inst{2} Saratov State University, 410012, Astrakhanskaya St. 83, Saratov,
  Russia\\
  \inst{3}R\&D Institute 'Volga', Saratov 410052, Russia }
\pacs{61.43.-j}{Disordered solids} \pacs{62.40.+i}{Anelasticity,
internal friction, stress relaxation, and mechanical resonances}
\pacs{63.20.kp}{Phonon–defect interactions}

\abstract{ We calculate the scattering rates of phonons on
two-level systems in disordered trigonal and hexagonal crystals.
We apply a model in which the two-level system, characterized by a
direction in space, is coupled to the strain field of the phonon
via a tensor of coupling constants. The structure of the tensor of
coupling constants is similar to the structure of the tensor of
elastic stiffness constants, in the sense that they are determined
by the same symmetry transformations. In this way, we emphasize
the anisotropy of the interaction of elastic waves with the
ensemble of two-level systems in disordered crystals. We also
point to the fact that the ratio $\gamma_l/\gamma_t$ has a much
broader range of allowed values in disordered crystals than in
isotropic solids.}

\begin{document}
\maketitle

\section{Introduction} \label{intro}

The ``universality'' of the glass-like properties of amorphous solids have been
pointed out almost four decades ago \cite{PhysRevB.4.2029.1971.Zeller}.
Some of these properties are: heat conductivity which is almost independent
of the chemical composition of the solid and proportional to $T^2$
(where $T$ is the temperature), specific heat proportional to $T$, and
a long-time heat release \cite{PhysRevB.4.2029.1971.Zeller,esquinazi:book,RevModPhys.74.991.2002.Pohl}.
All these properties are described theoretically with reasonable quantitative
accuracy by assuming that the amorphous solid contains dynamical defects
that can be described at low temperatures as an ensemble of two-level systems
(TLS) \cite{JLowTempPhys.7.351.1972.Philips,PhilMag.25.1.1972.Anderson}.
Nevertheless, glass-like properties have been found also in disordered crystals
\cite{SolidStateCommun19.335.1976.Lou,JPhysique.41.1193.1980.Doussineau,PhysRevB.42.5842.1990.Vanelstraete,PhysRevB.51.8158.1995.Laermans,PhysRevLett.81.3171.1998.Liu,Topp:thesis,PhysRevLett.75.1965.1995.Watson,ZPhysBCondensMatter.101.235.1996.Topp,PhysRevB.60.898.1999.Topp} and quasicrystals,\cite{PhysRevB.62.11437.2000.Thompson,PhysRevB.62.292.2000.Gianno,PhysRevLett.88.255901.2002.Bert}
only that in these materials they are not as universal as in amorphous
solids and, even more, they exhibit anisotropy.

The deep nature of the glass-like properties--and therefore of the ensemble
of the TLSs--remains elusive, despite
of the long and intensive efforts invested into their study.
This makes the study of disordered crystals especially interesting, since
there, knowing the structure of the unit cell and its modifications due
to disorder, we may know which are the tunneling entities and therefore
we may have additional information about the TLSs. Moreover, the observed
anisotropy of the glass-like properties, although unexplained,
represents additional information for the theoretical description, which
may help to improve the microscopic model.

In general, the thermal properties of a dielectric glass are
determined by the ensemble of TLSs, the phonon gas, and the interaction between
them. In the \textit{standard tunneling model} (STM) the TLS is
described in a basis that diagonalizes the interaction Hamiltonian between
the TLS and the phonon. In this basis, the Hamiltonian of the free
TLS and the interaction Hamiltonian are
\begin{eqnarray}
H_{TLS} = \frac{1}{2}\left(\begin{array}{cc}
\Delta & -\Lambda \\ -\Lambda & -\Delta
\end{array}\right)
%
\ {\rm and}\
%
H_{I} = \frac{1}{2}\left(\begin{array}{cc}
\delta & 0 \\ 0 & -\delta
\end{array}\right) ,&& \label{HTLS_HI}
\end{eqnarray}
respectively. The interaction element, $\delta$, is linear in the strain
field of the phonon, $[S]$, namely
\begin{equation}
\delta = 2\gamma_{ij}S_{ij} \label{delta_def0}
\end{equation}
where we assumed summation over the repeated subscripts. The symmetric second
rank tensor $[\gamma]$ characterizes the TLS and its ``deformability'' under
elastic strain.
For the convenience of the calculations we work
in the \textit{abbreviated subscript notations} and we write
$[S]$ and $[\gamma]$ as the six-dimensional vectors
$\bS=(S_{11},S_{22},S_{33},2S_{23},2S_{13},2S_{12})^t$ and
$\bgamma=(\gamma_{11},\gamma_{22},\gamma_{33},\gamma_{23},\gamma_{13},\gamma_{12})^t$, respectively (where the superscript $t$ denotes the transpose of a 
matrix or a vector).
To go further and obtain deeper information about the TLSs, in ref.
\cite{PhysRevB75.064202.2007.Anghel} $\bgamma$
was written as a product of two tensors: the first is a symmetric second rank
tensor describing the ``free'' TLS--we call it $\bT$ in abbreviated subscript
notations--and the second one is a fourth rank tensor, describing the coupling
between $\bT$ and $\bS$. In the abbreviated subscript notations, the
fourth rank tensor is a $6\times6$ matrix which we shall call $[R]$ and
represents the matrix of TLS-phonon coupling constants. So we
write eq. (\ref{delta_def0}) in matrix notations as
\cite{PhysRevB75.064202.2007.Anghel}
\begin{equation}
\delta = 2\bT^t\cdot[R]\cdot\bS . \label{delta_def1}
\end{equation}
The advantage of eq. (\ref{delta_def1}) is that it separates the tensor
$\bT$, which contains only the characteristics of the TLS, from the
matrix of coupling constants, $[R]$, in which are embedded the
characteristics of the interaction and has a general structure determined
by the symmetries of the lattice
\cite{PhysRevB75.064202.2007.Anghel,JPhysConfSer.92.12133.2007.Anghel}.
In general, the tensor $\bT$ was taken in the simple form,
$\bT=(t_1^2,t_2^2,t_3^2,2t_2t_3,2t_1t_3,2t_1t_2)^t$, where $t_1$, $t_2$, and
$t_3$ are the components of the unit vector $\hat\bt$, which determines the
``direction'' of the TLS. This simplification will be used also in this paper.
In the form (\ref{delta_def1}), the symmetries of the lattice are imposed
on the matrix $[R]$ by coordinates transformations that leave the lattice
invariant, whereas the distribution over the elements of $\bT$ is determined
by the distribution over the ``directions'', $\hat\bt$. Through the
properties of $[R]$, this
model predicts anisotropic glass properties of a crystal,
even for an ensemble of TLSs isotropically oriented.

In ref. \cite{PhysRevLett.2008.Anghel} we applied this model to
study the anisotropy of the glass properties in a disordered cubic
crystal and we compared our calculations with the experimental results of
Topp and coworkers \cite{Topp:thesis,ZPhysBCondensMatter.101.235.1996.Topp,PhysRevB.60.898.1999.Topp}. Unfortunately the experimental data published
to date is not enough to check the model of
ref. \cite{PhysRevB75.064202.2007.Anghel} or even to determine its
parameters. From the available data, in ref.
\cite{PhysRevB75.064202.2007.Anghel} we merely
obtained a relation between these parameters,
which  should be confirmed or not by future experiments.

In this paper we extend our calculations to two other classes of
crystal symmetries: trigonal \textit{32} and hexagonal.
The former symmetry class corresponds to (neutron-irradiated) quartz and
the latter to Na doped $\beta$-Al$_2$O$_3$. Both materials show glass-like
properties at low temperatures and strong anisotropy in the TLS-phonon
coupling \cite{PhysRevB.42.5842.1990.Vanelstraete,PhysRevB.51.8158.1995.Laermans,JPhysique.41.1193.1980.Doussineau}.

It is known that in isotropic amorphous materials the coupling
of TLSs with the phonon modes is described by the scalar coupling
constants, $\gamma_l$ and $\gamma_t$, obtained by
averaging the transition rates over the isotropic distribution of the
TLS orientations. In this way, from very general considerations, one gets
\cite{AnnNYAcadSci.279.173.1976.Halperin,SovPhysJETP.56.1334.1982.Gurevich,PhysRevB75.064202.2007.Anghel}
\begin{equation}
\left(\gamma_l/\gamma_t\right)^2 \ge 4/3 . \label{glgtiso}
\end{equation}
But in the model that we use here, this relation is affected by the
symmetry of the lattice and therefore it does not necessary hold in a
disordered crystal. This motivated us to discuss at 
the end of the next section 
the range of $\gamma_l/\gamma_t$ for a crystal with cubic symmetry.

\section{Phonon scattering rates in trigonal and hexagonal lattices}

\subsection{General considerations}

The transition amplitude from a quantum state consisting of an
unexcited TLS and $n_{\bk\sigma}+1$ phonons of wavevector $\bk$ and
polarization $\sigma$ ($\sigma=l,t$), $|n_{\bk\sigma}+1,\downarrow\rangle$,
into the state of $n_{\bk\sigma}$ phonons and excited TLS,
$|n_{\bk\sigma},\uparrow\rangle$, is
\begin{equation}
\langle n_{\bk\sigma},\uparrow|\tilde{H}_1|n_{\bk\sigma}+1,\downarrow\rangle
 = -\frac{\Lambda}{\epsilon}\sqrt{n_{\bk\sigma}}\bT^t\cdot[R]\cdot\bS_{\bk\sigma}
\label{eqn_matrix_element}
\end{equation}
where $\epsilon=\sqrt{\Delta^2+\Lambda^2}$ is the
\textit{excitation energy of the TLS}. Therefore the
phonon scattering rate by a TLS in its ground state is
\begin{equation}
\Gamma_{\bk\sigma}(\hat\bt) = \frac{2\pi}{\hbar}\frac{\Lambda^2n_{\bk\sigma}}
{\epsilon^2}|\bT^t\cdot[R]\cdot\bS_{\bk\sigma}|^2\delta(\epsilon-\hbar\omega).
\label{eqn_Gamma_bar}
\end{equation}
The main characteristic of the TLS-phonon interaction is contained in the
quantity $M_{\bk,\sigma}(\hat\bt)\equiv\bT^t\cdot[R]\cdot\bS_{\bk\sigma}$.
As explained in the Introduction, the TLS-phonon interaction bear an
intrinsic anisotropy through the matrix $[R]$, on which the symmetries of
the lattice are imposed.
To calculate the
average scattering rate of a phonon by the ensemble of TLSs, we have to
average over the distribution of $\hat\bt$. To reduce the number of
degrees of freedom of the problem, in what follows we shall assume that
$\hat\bt$ is isotropically oriented.

\subsection{Trigonal lattice}

For a trigonal lattice of symmetry class \textit{32} (the symmetry of
quartz), the matrix $[R]$ has the form \cite{Auld:book}
\begin{eqnarray}
[R] = \left(\begin{array}{cccccc}
r_{11}&r_{12}&r_{13}&r_{14}&0&0\\
r_{12}&r_{11}&r_{13}&-r_{14}&0&0\\
r_{13}&r_{13}&r_{33}&0&0&0\\
r_{14}&-r_{14}&0&r_{44}&0&0\\
0&0&0&0&r_{44}&r_{14}\\
0&0&0&0&r_{14}&\frac{r_{11}-r_{12}}{2}
\end{array}\right), && \label{R_trigonal}
\end{eqnarray}
similar to that of the tensor of elastic stiffness constants, $[c]$,
with $c_{ij}$ replaced by $r_{ij}$ \cite{PhysRevB75.064202.2007.Anghel,JPhysConfSer.92.12133.2007.Anghel}.
The system of coordinates that we use here is such that the $z$ and $x$
axes are the 3-fold and 2-fold rotational symmetry axes, respectively,
while the $y$ axis is perpendicular to both $x$ and $z$. Solving the
Christoffel equation we find that the crystal can sustain pure longitudinal
waves propagating  along the $x$ and $z$ axes, and pure transversal waves
propagating along the $y$ and $z$ axes. The sound velocities of the
longitudinal waves propagating in the $x$ and $z$ directions are
$v_{\hat\bx,l}=\sqrt{c_{11}/\rho}$ and $v_{\hat\bz,l}=\sqrt{c_{33}/\rho}$,
respectively,
where $\rho$ is the density of the material. The transversal waves
propagating in the $z$ direction have a sound velocity,
$v_{\hat\bz,t}=\sqrt{c_{44}/\rho}$, independent of the polarisation direction.
The pure transversal waves propagating in the $x$ direction should be
polarized only in the $z$ direction and have a sound velocity 
$v_{\hat\bx,t}=\sqrt{(c_{11}-c_{12})/2\rho}$--transversal waves 
polarized in other directions are not eigenvectors of the Christoffel 
equation. If we define the
direction $\hat\bt$ by the two Euler angles $\theta$ and $\phi$, 
$\hat\bt=(\sin\theta\cos\phi,\sin\theta\sin\phi,\cos\theta)^t$, then for 
the longitudinal waves propagating
in the $\hat\bx$ and $\hat\bz$ directions we have
\begin{subequations} \label{Ms}
\begin{eqnarray}
M_{k\hat\bx,l}(\theta,\phi) &=& ik[r_{11}\sin^2(\theta)\cos^2(\phi)+r_{12}\sin^2(\theta)\sin^2(\phi) \nonumber \\
&& +r_{13}\cos^2(\theta)+r_{14}\sin(2\theta)\sin(\phi)]
\label{Mzl}
\end{eqnarray}
and
\begin{equation}
M_{k\hat\bz,l}(\theta,\phi) = ik[r_{13}\sin^2(\theta)+r_{33}\cos^2(\theta)],
\label{Mxl}
\end{equation}
respectively, whereas for the transversely polarized waves propagating in the
$\hat\by$ and $\hat\bz$ directions we have
\begin{eqnarray}
M_{k\hat\by,t}(\theta,\phi) &=& ik\sin(\theta)[r_{14}\sin(\theta)\cos(2\phi)
\nonumber \\
&& +2r_{44}\cos(\theta)\sin(\phi)] \label{Myzt}
\end{eqnarray}
and
\begin{eqnarray}
M_{k\hat\bz,t}(\theta,\phi) &=& 2ik\sin(\theta)\cos(\phi)[r_{44}\cos(\theta)
\nonumber \\
&& +r_{14}\sin(\theta)\sin(\phi)), \label{Mzxt}
\end{eqnarray}
\end{subequations}
respectively. For the transversal wave propagating in the $\hat\bz$ direction
we choose the polarization along the $x$ axis--this choice becomes
irrelevant after averaging over the directions $\hat\bt$, which we shall do
next.

The phonon absorption rates are calculated by averaging
(\ref{eqn_Gamma_bar}) over the ensemble of TLSs. If we denote
by $f(\theta,\phi)$ the distribution over the angles of $\hat\bt(\theta,\phi)$,
then we have
\begin{eqnarray}
\tau^{-1}_{\bk\sigma} &=& \frac{P_0\tanh\left(\frac{\epsilon}{2\kbt}
\right)}{2\hbar}n_{\bk\sigma}\int_0^\pi d\theta
\int_0^{2\pi}d\phi\sin\theta
 \nonumber \\
&& \times|M_{\bk\sigma}[\hat\bt(\theta,\phi)]|^2 f(\theta,\phi) \equiv 
\frac{2\pi P_0\tanh\left(\frac{\epsilon}{2\kbt}\right)}{\hbar} \nonumber \\
&& \times n_{\bk\sigma}\langle|M_{\bk\sigma}(\hat\bt)|^2\rangle.
\label{av_def}
\end{eqnarray}
As mentioned before, we shall use the isotropic distribution,
$f(\theta,\phi)=1$.
Plugging eqs. (\ref{Ms}) one by one into (\ref{av_def}), we get the
scattering rates
\begin{subequations} \label{taus}
\begin{eqnarray}
\tau^{-1}_{k\hat\bx,l} &=&
\frac{1}{15}(3r_{11}^{2}+2r_{11}r_{12}+2r_{13}r_{11}+3r_{12}^{2}+2r_{12}r_{13}
\nonumber \\
&&+3r_{13}^{2}+4r_{14}^{2})\cdot \frac{2\pi P_0N^2nk^2}{\hbar}
\tanh\left(\frac{\epsilon}{2\kbt}\right),
\label{tauXl} \\
\tau^{-1}_{k\hat\bz,l} &=&
\frac{8r_{13}^{2}+4r_{13}r_{33}+3r_{33}^{2}}{15}\cdot\frac{2\pi P_0N^2nk^2}
{\hbar} \nonumber \\
&& \times\tanh\left(\frac{\epsilon}{2\kbt}\right) ,
\label{tauZl} \\
\tau^{-1}_{k\hat\by,t} &=&
\frac{4(r_{14}^2+r_{44}^2)}{15}
\cdot\frac{2\pi P_0N^2nk^2}{\hbar}\tanh\left(\frac{\epsilon}{2\kbt}\right),
\label{tauYZt} \\
\tau^{-1}_{k\hat\bz,t} &=&
\frac{4(4r_{14}^2+r_{44}^2)}{15}
\cdot\frac{2\pi P_0N^2nk^2}{\hbar}\tanh\left(\frac{\epsilon}{2\kbt}\right) ,
\label{tauZt}
\end{eqnarray}
\end{subequations}
where by $N$ we denoted the normalization constant of the
phonon ($N=\sqrt{\hbar/(2V\rho\omega)}$) and by $n$ the thermal population
of the phonon mode ($n=[\exp(\beta\hbar\omega)-1]^{-1}$); $V$ is the volume
of the solid.
Comparing eqs. (\ref{taus}) with the standard formula for the phonon
scattering rates,
\begin{equation}
\left(\tau^{\rm (STM)}_{\bk,\sigma}\right)^{-1}
= \gamma^2_{\hat\bk,\sigma}\frac{2\pi P_0N^2nk^2}{\hbar}
\tanh\left(\frac{\epsilon}{2\kbt}\right),
\label{av_STM}
\end{equation}
we obtain the anisotropic values of the $\gamma$ parameters,
\begin{subequations} \label{gammas}
\begin{eqnarray}
\gamma^2_{k\hat\bx,l} &=&
\frac{2(r_{11}^{2}+r_{12}^{2}+r_{13}^{2})+(r_{11}+r_{12}+r_{13})^2+4r_{14}^{2}}{15},
\nonumber \\ && \label{gammaXl} \\
\gamma^2_{k\hat\bz,l} &=&
\frac{8r_{13}^{2}+4r_{13}r_{33}+3r_{33}^{2}}{15} , \label{gammaZl} \\
\gamma^2_{k\hat\by,t} &=&
\frac{4(r_{14}^2+r_{44}^2)}{15}, \label{gammaYZt} \\
\gamma^2_{k\hat\bz,t} &=&
\frac{4(4r_{14}^2+r_{44}^2)}{15} . \label{gammaZt}
\end{eqnarray}
\end{subequations}

\subsection{Hexagonal lattice}

The difference between the trigonal lattice of symmetry \textit{32} and
the hexagonal lattice is that $r_{14}$ and $c_{14}$ are zero. This
enhancement of symmetry allows propagation of pure longitudinal and
transversal waves in all the three directions, $x$, $y$, and $z$. The
sound velocities of the longitudinal waves in these three directions
are $\sqrt{c_{11}/\rho}$ for $x$ and $y$ directions, and
$\sqrt{c_{33}/\rho}$ for the $z$ direction. For the transversal waves
propagating in the $x$ direction, the ones polarized in the $y$
direction propagate with the velocity $\sqrt{(c_{11}-c_{12})/2\rho}$
whereas the ones polarized in the $z$ direction propagate with the
velocity $\sqrt{c_{44}/\rho}$. The transversal waves propagating in the
$y$ direction are similar to the ones propagating in the $x$ direction:
the waves polarized in the $x$ direction have a sound velocity of
$\sqrt{(c_{11}-c_{12})/2\rho}$, whereas the ones polarized in the $z$
direction have a sound velocity of $\sqrt{c_{44}/\rho}$. Finally, the
transversal waves propagating in the $z$ direction have all the same
sound velocity, $\sqrt{c_{44}/\rho}$.

For the quantities $M$, we get
\begin{subequations} \label{MsH}
\begin{eqnarray}
M_{k\hat\bx,l}(\theta,\phi) &=&
ik[r_{11}\sin^2(\theta)\cos^2(\phi)+r_{12}\sin^2(\theta)\sin^2(\phi)
\nonumber \\ && +r_{13}\cos^2(\theta)], \label{MxxH} \\
M_{k\hat\bx,\hat\by,t}(\theta,\phi) &=& ik\sin^2(\theta)\sin(2\phi)\frac{r_{11}-r_{12}}{2},
\label{MxyH} \\
M_{k\hat\bx,\hat\bz,t}(\theta,\phi) &=& ik\sin(2\theta)\cos(\phi)r_{44},
\label{MxzH} \\
M_{k\hat\by,l}(\theta,\phi) &=& ik[r_{11}\sin^2(\theta)\sin^2(\phi)+r_{12}\sin^2(\theta)\cos^2(\phi) \nonumber \\ && +r_{13}\cos^2(\theta)],
\label{MyyH} \\
M_{k\hat\by,\hat\bx,t}(\theta,\phi) &=& ik\sin^2(\theta)\sin(2\phi)\frac{r_{11}-r_{12}}{2},
\label{MyxH} \\
M_{k\hat\by,\hat\bz,t}(\theta,\phi) &=& ik\sin(2\theta)\sin(\phi)r_{44},
\label{MyzH} \\
M_{k\hat\bz,l}(\theta,\phi) &=& ik(r_{13}\sin^2(\theta)+r_{33}\cos^2(\theta)),
\label{MzzH} \\
M_{k\hat\bz,\hat\bx,t}(\theta,\phi) &=& ik\sin(2\theta)\cos(\phi)r_{44},
\label{MzxH} \\
M_{k\hat\bz,\hat\by,t}(\theta,\phi) &=& ik\sin(2\theta)\sin(\phi)r_{44},
\label{MzyH}
\end{eqnarray}
\end{subequations}
in obvious notations: the first subscript indicates the propagation
direction while the second one is used only for transversal waves and
denotes the direction of polarization. We plug
these formulae into eq. (\ref{av_def}) to get
\begin{equation}
\left(\tau^{\rm (H)}_{\bk,\sigma}\right)^{-1}
= (\gamma^{\rm (H)}_{\hat\bk,\sigma})^2\frac{2\pi P_0N^2nk^2}{\hbar}
\tanh\left(\frac{\epsilon}{2\kbt}\right),
\label{av_H}
\end{equation}
where the superscript $(H)$ stands for \textit{hexagonal} and is used to make
the difference between these quantities
and the ones calculated in the preceding subsection. As before, we get
the $\gamma$ constants:
\begin{subequations} \label{gammasH}
\begin{eqnarray}
(\gamma^{\rm (H)}_{k\hat\bx,l})^2  &=&
\frac{2(r_{11}^{2}+r_{12}^{2}+r_{13}^{2})+(r_{11}+r_{12}+r_{13})^2}{15},
\nonumber \\ &=& (\gamma^{\rm (H)}_{k\hat\by,l})^2 \label{gammaXlH} \\
(\gamma^{\rm (H)}_{k\hat\bz,l})^2 &=&
\frac{8r_{13}^{2}+4r_{13}r_{33}+3r_{33}^{2}}{15} , \label{gammaZlH} \\
(\gamma^{\rm (H)}_{k\hat\by,\hat\bx,t})^2 &=& (\gamma^{\rm (H)}_{k\hat\bx,\hat\by,t})^2
= \frac{(r_{11}-r_{12})^2}{15}, \label{gammaYXtH} \\
(\gamma^{\rm (H)}_{k\hat\bx,\hat\bz,t})^2 &=& (\gamma^{\rm (H)}_{k\hat\by,\hat\bz,t})^2
= (\gamma^{\rm (H)}_{k\hat\bz,\hat\bx,t})^2 = (\gamma^{\rm (H)}_{k\hat\bz,\hat\by,t})^2
= \frac{4r_{44}^2}{15} . \nonumber \\ && \label{gammaXZtH}
\end{eqnarray}
\end{subequations}
We notice that
the constants
$\gamma^{\rm (H)}_{k\hat\bx,l}$,
$\gamma^{\rm (H)}_{k\hat\bz,l}$, $\gamma^{\rm (H)}_{k\hat\by,t}$ and
$\gamma^{\rm (H)}_{k\hat\bz,t}$ are equal to $\gamma_{k\hat\bx,l}$,
$\gamma_{k\hat\bz,l}$, $\gamma_{k\hat\by,t}$ and $\gamma_{k\hat\bz,t}$,
respectively (eqs. \ref{gammas}) if in the last ones we set $r_{14}=0$.

\subsection{Range of $\gamma_l/\gamma_t$ \label{Sglgtcube}}

We notice also that in general the relation (\ref{glgtiso}), valid for
isotropic mediums, is not necessarily valid
for crystals, which have lower symmetry. For the lattices studied
above, the ratio $\gamma^2_l/\gamma^2_t$ in any of the three directions
has complicated expressions in terms of the components of $[R]$.
For the trigonal lattice
$[R]$ has six independent components, whereas for the hexagonal lattice
it has five. Therefore a discussion about the ranges of $\gamma^2_l/\gamma^2_t$
for such symmetries would be too general to be of much use.

The simplest lattice we can discuss is the cubic lattice; its $[c]$ and
$[R]$ matrices have only three independent components:
$c_{11}$, $c_{12}$, and $c_{44}$ for $[c]$ and $r_{11}$, $r_{12}$, and
$r_{44}$ for $[R]$.
The $\gamma_l^{\rm (c)}$ and $\gamma_t^{\rm (c)}$ (we use the
superscript $(c)$ to refer to the cubic lattice) constants
have been calculated in Ref. \cite{PhysRevLett.2008.Anghel}
for longitudinal and transversal waves propagating in the
$\langle100\rangle$, $\langle110\rangle$, and $\langle111\rangle$
crystallographic directions and for an isotropic distribution of
TLS orientations. Using the results of Ref. \cite{PhysRevLett.2008.Anghel},
we calculate the ratios $(\gamma_l^{\rm (c)}/\gamma_t^{\rm (c)})^2$ for the
waves propagating in the three directions mentioned above. Denoting
$\zeta\equiv r_{12}/r_{11}$ and $\xi\equiv r_{44}/r_{11}$, we obtain
\begin{subequations} \label{gl_gt_c}
\begin{eqnarray}
\left(\frac{\gamma_l^{\rm (c)}}{\gamma_t^{\rm (c)}}\right)^2_{\langle100\rangle}
&=& \frac{3+4\zeta+8\zeta^2}{4\xi^2} \label{gl_gt_c100} \\
\left(\frac{\gamma_l^{\rm (c)}}{\gamma_t^{\rm (c)}}\right)^2_{\langle110\rangle,1}
&=& \frac{2+6\zeta+7\zeta^2+4\xi^2}{4\xi^2} \label{gl_gt_c1101} \\
\left(\frac{\gamma_l^{\rm (c)}}{\gamma_t^{\rm (c)}}\right)^2_{\langle110\rangle,2}
&=& \frac{2+6\zeta+7\zeta^2+4\xi^2}{(1-\zeta)^2} \label{gl_gt_c1102} \\
\left(\frac{\gamma_l^{\rm (c)}}{\gamma_t^{\rm (c)}}\right)^2_{\langle111\rangle}
&=& \frac{5+20\zeta+20\zeta^2+16\xi^2}{2[(1-\zeta)^2+2\xi^2]}\label{gl_gt_c111}
\end{eqnarray}
\end{subequations}
Note that in the $\langle110\rangle$ direction there are two transversal
elastic waves, of reciprocally perpendicular polarization, propagating
with different sound velocities.
The isotropy condition for $[R]$ is $\zeta+2\xi=1$, which sets the
range of $(\gamma_l^{\rm (c)}/\gamma_t^{\rm (c)})^2$ to $[4/3,\infty)$,
as stated in eq. (\ref{glgtiso}). If the lattice has lower symmetry,
then $\zeta+2\xi\ne1$ and we introduce the parameter $K$ to
quantify the anisotropy, by imposing $\zeta+2K\xi=1$; therefore
$K=1$ corresponds to the isotropic case. We calculate the
dependence on $K$ of the ranges of $(\gamma_l^{\rm (c)}/\gamma_t^{\rm (c)})^2$
in the three crystallographic directions of eqs. (\ref{gl_gt_c}).
Replacing $\zeta$ by $1-2K\xi$ into (\ref{gl_gt_c}), we get
\begin{subequations} \label{gl_gt_cK}
\begin{eqnarray}
\left(\frac{\gamma_l^{\rm (c)}}{\gamma_t^{\rm (c)}}\right)^2_{\langle100\rangle}
&=& \frac{15}{4}\cdot\frac{1}{\xi^2}-\frac{10K}{\xi}+8K^2 ,
\label{gl_gt_c100K} \\
\left(\frac{\gamma_l^{\rm (c)}}{\gamma_t^{\rm (c)}}\right)^2_{\langle110\rangle,1}
&=& \frac{15}{4}\cdot\frac{1}{\xi^2}-\frac{10K}{\xi}+7K^2+1 ,
\label{gl_gt_c1101K} \\
\left(\frac{\gamma_l^{\rm (c)}}{\gamma_t^{\rm (c)}}\right)^2_{\langle110\rangle,2}
&=& \frac{15}{4K^2}\cdot\frac{1}{\xi^2}-\frac{10}{K\xi}+7+\frac{1}{K^2} ,
\label{gl_gt_c1102K} \\
\left(\frac{\gamma_l^{\rm (c)}}{\gamma_t^{\rm (c)}}\right)^2_{\langle111\rangle}
&=& \frac{45}{4(2K^2+1)}\cdot\frac{1}{\xi^2}-\frac{30K}{2K^2+1}\cdot
\frac{1}{\xi} \nonumber \\
&&  + \frac{4(5K^2+1)}{2K^2+1} \label{gl_gt_c111K} .
\end{eqnarray}
\end{subequations}
Obviously, the condition $K=1$ restores the isotropic equation for
$(\gamma_l/\gamma_t)^2$ \cite{PhysRevB75.064202.2007.Anghel}.
What is interesting to note is that all the equations (\ref{gl_gt_cK})
are quadratic in $1/\xi$ and attain their minima at
\begin{equation}
\left.\frac{1}{\xi}\right|_{\tx{min}} = \frac{4K}{3} ,
\end{equation}
and we obtain the following constraints on $(\gamma_l/\gamma_t)^2$
in the three propagation directions of the cubic crystal:
\begin{subequations} \label{gl_gt_cKmin}
\begin{eqnarray}
\left(\frac{\gamma_l^{\rm (c)}}{\gamma_t^{\rm (c)}}\right)^2_{\langle100\rangle}
&\ge& \frac{4K^2}{3} ,
\label{gl_gt_c100Kmin} \\
\left(\frac{\gamma_l^{\rm (c)}}{\gamma_t^{\rm (c)}}\right)^2_{\langle110\rangle,1}
&\ge& \frac{K^2+3}{3} ,
\label{gl_gt_c1101Kmin} \\
\left(\frac{\gamma_l^{\rm (c)}}{\gamma_t^{\rm (c)}}\right)^2_{\langle110\rangle,2}
&\ge& \frac{K^2+3}{3K^2} ,
\label{gl_gt_c1102Kmin} \\
\left(\frac{\gamma_l^{\rm (c)}}{\gamma_t^{\rm (c)}}\right)^2_{\langle111\rangle}
&\ge& \frac{4}{2K^2+1} \label{gl_gt_c111Kmin} .
\end{eqnarray}
\end{subequations}
Now we can see that although for $K=1$ all the conditions become
identical, namely $(\gamma_l^{\rm (c)}/\gamma_t^{\rm (c)})^2\ge4/3$, for
$K\ne1$ the lower limits of $(\gamma_l^{\rm (c)}/\gamma_t^{\rm (c)})^2$
vary differently. For example for $K\gg1$, the lower limits for
$(\gamma_l^{\rm (c)}/\gamma_t^{\rm (c)})^2_{\langle100\rangle}$ and
$(\gamma_l^{\rm (c)}/\gamma_t^{\rm (c)})^2_{\langle110\rangle,1}$ become
very big ($\propto K^2$), the lower limit of
$(\gamma_l^{\rm (c)}/\gamma_t^{\rm (c)})^2_{\langle110\rangle,2}$
converges to $1/3$, whereas the lower limit of
$(\gamma_l^{\rm (c)}/\gamma_t^{\rm (c)})^2_{\langle111\rangle}$ converges to
zero.

If $K\ll1$, the situation is the other way around. The limit
(\ref{gl_gt_c100Kmin}) converges to zero, (\ref{gl_gt_c1101Kmin}) and
(\ref{gl_gt_c111Kmin}) to 1 and 4, respectively, whereas
the limit value (\ref{gl_gt_c1102Kmin}) converges to infinity, like
$1/K^2$. Therefore in a cubic crystal,
$\gamma_l^{\rm (c)}$ can become smaller $\gamma_t^{\rm (c)}$ if the
matrix $[R]$ deviates significantly from the isotropic condition.

\section{Conclusions}

We calculated the average phonon scattering rates on TLS in trigonal and
hexagonal crystals, to emphasize the anisotropy imposed
by the lattice symmetry. 
The parameters of the model may be obtained by measuring 
$\gamma_l$ and/or $\gamma_t$ in some crystallographic directions
and this enables one to calculate the coupling of TLSs with
phonons propagating in any other direction.
The number of $\gamma_l$s and $\gamma_t$s needed, depends on the number of 
independent parameters of the tensor of coupling constants, $[R]$, 
which is determined by the symmetry of the lattice. 

We showed that the allowed
limits of the ratio $\gamma^2_l/\gamma^2_t$ in crystals with different
symmetries are different from the one imposed in isotropic materials,
which is $\gamma^2_l/\gamma^2_t\ge 4/3$. In principle
$\gamma^2_l/\gamma^2_t$ in crystals may take any value.

The calculations can be extended easily to disordered crystals of any 
symmetry. Moreover, although we used in our calculations an isotropic 
distribution over the TLS orientations, the comparison of our calculations 
with experimental data would enable one to find if our assumption is true 
or not. If it is not true, 
one can determine, at least in principle, the distribution over 
the orientations of the TLSs.

\acknowledgments

We are grateful to Prof. R. Pohl, Prof. K. A. Topp for very useful and
motivating correspondence.
This work was partially supported by the NATO grant, EAP.RIG 982080.


\begin{thebibliography}{10}

\bibitem{PhysRevB.4.2029.1971.Zeller}
R.~C. Zeller and R.~O. Pohl.
\newblock {\em Phys. Rev. B}, 4:2029, 1971.

\bibitem{esquinazi:book}
P.~Esquinazi.
\newblock {\em Tunneling systems in amorphous and crystalline solids}.
\newblock Springer, 1998.

\bibitem{RevModPhys.74.991.2002.Pohl}
Robert~O. Pohl, Xiao Liu, and EunJoo Thompson.
\newblock {\em Rev. Mod. Phys.}, 74:991, 2002.

\bibitem{JLowTempPhys.7.351.1972.Philips}
W.~A. Philips.
\newblock {\em J. Low Temp. Phys.}, 7:351, 1972.

\bibitem{PhilMag.25.1.1972.Anderson}
P.~W. Anderson, B.~I. Halperin, and C.~M. Varma.
\newblock {\em Phil. Mag.}, 25:1, 1972.

\bibitem{SolidStateCommun19.335.1976.Lou}
L.~F. Lou.
\newblock {\em Solid State Commun.}, 19:335, 1976.

\bibitem{JPhysique.41.1193.1980.Doussineau}
P.~Doussineau, C.~Fr{\'e}nois, R.~G. Leisure, A.~Levelut, and J.-Y. Prieur.
\newblock {\em J. Physique}, 41:1193, 1980.

\bibitem{PhysRevB.42.5842.1990.Vanelstraete}
A.~Vanelstraete and C.~Laermans.
\newblock {\em Phys. Rev. B}, 42:5842, 1990.

\bibitem{PhysRevB.51.8158.1995.Laermans}
C.~Laermans and V.~Keppens.
\newblock {\em Phys. Rev. B}, 51:8158, 1995.

\bibitem{PhysRevLett.81.3171.1998.Liu}
Xiao Liu, P.~D. Vu, R.~O. Pohl, F.~Schiettekatte, and S.~Roorda.
\newblock {\em Phys. Rev. Lett.}, 81:3171, 1998.

\bibitem{Topp:thesis}
K.~A. Topp.
\newblock {\em Effects of random strains on tunneling states in crystals}.
\newblock PhD thesis, 1997.

\bibitem{PhysRevLett.75.1965.1995.Watson}
Susan~K. Watson.
\newblock {\em Phys. Rev. Lett.}, 75:1965, 1995.

\bibitem{ZPhysBCondensMatter.101.235.1996.Topp}
K.~A. Topp and D.~G. Cahill.
\newblock {\em Z. Phys. B: Condens. Matter}, 101:235, 1996.

\bibitem{PhysRevB.60.898.1999.Topp}
Karen~A. Topp, EunJoo Thompson, and R.~O. Pohl.
\newblock {\em Phys. Rev. B}, 60:898, 1999.

\bibitem{PhysRevB.62.11437.2000.Thompson}
EunJoo Thompson, P.~D. Vu, and R.~O. Pohl.
\newblock {\em Phys. Rev. B}, 62:11437, 2000.

\bibitem{PhysRevB.62.292.2000.Gianno}
K.~Giann{\`o}, A.~V. Sologubenko, M.~A. Chernikov, H.~R. Ott, I.~R. Fisher, and
  P.~C. Canfield.
\newblock {\em Phys. Rev. B}, 62:292, 2000.

\bibitem{PhysRevLett.88.255901.2002.Bert}
F.~Bert, G.~Bellessa, and B.~Grushko.
\newblock {\em Phys. Rev. Lett.}, 88:255901, 2002.

\bibitem{PhysRevB75.064202.2007.Anghel}
D.~V. Anghel, T.~K\"uhn, Y.~M. Galperin, and M.~Manninen.
\newblock {\em Phys. Rev. B}, 75:064202, 2007.

\bibitem{JPhysConfSer.92.12133.2007.Anghel}
D.~V. Anghel, T.~K{\"u}hn, Y.~M. Galperin, and M~Manninen.
\newblock {\em J. Phys.: Conf. Series}, 92:012133, 2007.

\bibitem{PhysRevLett.2008.Anghel}
D.~V. Anghel and D.~V. Churochkin.
\newblock arXiv:0804.1481, 2008.

\bibitem{AnnNYAcadSci.279.173.1976.Halperin}
B.~I. Halperin.
\newblock {\em Ann. N.Y. Acad. Sci.}, 279:173, 1976.

\bibitem{SovPhysJETP.56.1334.1982.Gurevich}
V.~L. Gurevich and D.~A. Parshin.
\newblock {\em Sov. Phys. JETP}, 56:1334, 1982.

\bibitem{Auld:book}
B.~A. Auld.
\newblock {\em Acoustic Fields and Waves in Solids, 2nd Ed.}
\newblock Robert E. Krieger Publishing Company, 1990.

\end{thebibliography}

\end{document}